# Any $J$-state solution of the DKP equation for a vector deformed Woods-Saxon potential

M. Hamzavi[1*], S. M. Ikhdair[2]

[1]*Department of Basic Sciences, Shahrood Branch, Islamic Azad University, Shahrood, Iran*

[2]*Physics Department, Near East University, Nicosia, North Cyprus, Mersin 10, Turkey*

[*]*Corresponding author: Tel.:+98 273 3395270, fax: +98 273 3395270*

Email: majid.hamzavi@gmail.com

**Abstract**

By using the Pekeris approximation, the Duffin–Kemmer–Petiau (DKP) equation is investigated for a vector deformed Woods-Saxon (dWS) potential. The parametric Nikiforov-Uvarov (NU) method is used in calculations. The approximate energy eigenvalue equation and the corresponding wave function spinor components are calculated for any total angular momentum $J$ in closed form. The exact energy equation and wave function spinor components are also given for the $J = 0$ case. We use a set of parameter values to obtain the numerical values for the energy states with various values of quantum levels $(n, J)$ and potential's deformation constant $q$ and width $R$.

## 1. Introduction

The first-order DKP formalism describes spin-0 and spin-1 particles has been used to analyze relativistic interactions of spin-0 and spin-1 hadrons with nuclei as an alternative to their conventional second-order Klein-Gordon (KG) and Proca equations [1-5]. The DKP equation is a direct generalization to the Dirac particles of integer spin in which one replaces the gamma matrices by beta metrics but verifying a more complicated algebra as DKP algebra [6-14]. Fainberg and Pimentel presented a strict proof of equivalence between DKP and Klein-Gordon theories for physical S-matrix elements in the case of charged scalar particles interacting in minimal way with an external or quantized electromagnetic field [15-16]. Boutabia-Chéraitia and Boudjedaa solved the DKP equation in the presence of Woods–Saxon potential for spin 1 and spin 0 and they also deduced the transmission and reflection coefficients [17]. Kulikov et al. offered a new oscillator model with different form of the nonminimal substitution within the framework of the DKP equation [18]. Yaşuk et al. presented an application of the relativistic DKP equation in the presence of a vector deformed Hulthén potential for spin zero particles by using the Nikiforov-Uvarov (NU) method [19]. Boztosun et al. presented a simple exact analytical solution of the relativistic DKP equation within the framework of the asymptotic iteration method and determined exact bound state energy eigenvalues and corresponding eigenfunctions for the relativistic harmonic oscillator as well as the Coulomb potentials [20]. Kasri and Chetouani determined the bound state energy eigenvalues for the relativistic DKP oscillator and DKP Coulomb potentials by using an exact quantization rule [21]. de Castro explored the problem of spin-0 and spin-1 bosons subject to a general mixing of minimal and nonminimal vector cusp potentials in a unified way in the context of the DKP theory [22]. Chargui et al. solved the DKP equation with a pseudoscalar linear plus Coulomb-like potential in a two-dimensional space–time [23].

The deformed Woods-Saxon (dWS) potential is a short range potential and widely used in nuclear, particle, atomic, condensed matter and chemical physics [24-30]. This potential is reasonable for nuclear shell models and used to represent the distribution of nuclear densities. The dWS and spin-orbit interaction are important and applicable to deformed nuclei [31] and to strongly deformed nuclides [32]. The dWS potential parameterization at large deformations for plutonium $^{237,239,241}Pu$ odd isotopes was analyzed [26]. The structure of single-particle states in the second minima of $^{237,239,241}Pu$ has been calculated with an exactly dWS potential. The Nuclear shape was parameterized. The parameterization of the spin-orbit part of the potential was obtained in the region corresponding to large deformations (second minima) depending only on the nuclear surface area. The spin-orbit interaction of a particle in a non-central self consistent field of the WS type potential was investigated for light nuclei and the scheme of single-particle states has been found for mass number $A_0 = 10$ and 25 [31]. Two parameters of the spin-orbit part of the dWS potential, namely the strength parameter and radius parameter were adjusted to reproduce the spins for the values of the nuclear deformation parameters [33].

Therefore, it would be interesting and important to solve the DKP equation for the dWS since it has extensively used to describe the bound and scattering states of the above mentioned interesting deformed nuclear models.

In this paper, we will study the DKP equation with a vector dWS potential for non-zero total angular momentum, i.e. $J \neq 0$. Under these conditions, the DKP equation has no exact solutions and we use the Pekeris approximation.

This work is arranged as follows: in section 2, the DKP formalism is given briefly and discussed under a vector potential. In section 3, parametric generalization of the NU method is introduced. We solve the DKP equation with a vector dWS potential in section 4. We also obtain numerical energy results for any arbitrary $(n,J)$ state with various values of potential's deformation constant $q$ and width $R$. The exact energy equation and wave function spinor components are also given for the $J=0$ case in section 4. Finally, section 5 is for our conclusion.

## 2. Review of the DKP formalism

The first order relativistic DKP equation for a free spin-0 or spin-1 particle of mass $m$ is

$$(i\beta^{\mu}\partial_{\mu} - m)\psi_{DKP} = 0, \quad (1)$$

where $\beta^{\mu}$ ($\mu = 0,1,2,3$) matrices satisfy the commutation relation

$$\beta^{\mu}\beta^{\nu}\beta^{\lambda} + \beta^{\lambda}\beta^{\nu}\beta^{\mu} = g^{\mu\nu}\beta^{\lambda} + g^{\nu\lambda}\beta^{\mu}, \quad (2)$$

which defines the so-called DKP algebra. The algebra generated by the $4\beta^{N}$ 's has three irreducible representations: a ten dimensional one that is related to $S=1$, a five dimensional one relevant for $S=0$ (spinless particles) and a one dimensional one which is trivial. In the spin-0 representation, $\beta^{\mu}$ are $5\times 5$ matrices defined as ($i=1,2,3$)

$$\beta^{0} = \begin{pmatrix} \theta & \tilde{0} \\ \bar{0}_{T} & 0 \end{pmatrix}, \quad \beta^{i} = \begin{pmatrix} \tilde{0} & \rho^{i} \\ -\rho^{i}_{T} & 0 \end{pmatrix}, \quad (3)$$

with $\tilde{0}$, $\bar{0}$, $0$ as $2\times 2$, $2\times 2$, $3\times 3$ zero matrices, respectively, and

$$\theta=\begin{pmatrix}0 & 1\\ 1 & 0\end{pmatrix},\ \rho^1=\begin{pmatrix}-1 & 0 & 0\\ 0 & 0 & 0\end{pmatrix},\ \rho^2=\begin{pmatrix}0 & -1 & 0\\ 0 & 0 & 0\end{pmatrix},\ \rho^3=\begin{pmatrix}0 & 0 & -1\\ 0 & 0 & 0\end{pmatrix}. \tag{4}$$

For spin one particle, $\beta^{\mu}$ are $10\times 10$ matrices given by

$$\beta^0=\begin{pmatrix}0 & \bar{0} & \bar{0} & \bar{0}\\ \bar{0}^T & 0 & I & I\\ \bar{0}^T & I & 0 & 0\\ \bar{0}^T & 0 & 0 & 0\end{pmatrix},\qquad \beta^i=\begin{pmatrix}0 & \bar{0} & e_i & \bar{0}\\ \bar{0}^T & 0 & 0 & -is_i\\ e_i^T & 0 & 0 & 0\\ \bar{0}^T & -is_i & 0 & 0\end{pmatrix}, \tag{5}$$

where $s_i$ are the usual $3\times 3$ spin one matrices

$$\bar{0}=\begin{pmatrix}0 & 0 & 0\end{pmatrix},\ e_1=\begin{pmatrix}1 & 0 & 0\end{pmatrix},\ e_2=\begin{pmatrix}0 & 1 & 0\end{pmatrix},\ e_3=\begin{pmatrix}0 & 0 & 1\end{pmatrix}, \tag{6}$$

$I$ and $0$ are the identity and zero matrices, respectively. While the dynamical state $\psi_{DKP}$ is a five component spinor for spin zero particles, it has ten component spinors for $S=1$ particles. The solution of the DKP equation for a particle in a central field needs consideration since earlier work [11]. It is convenient to recall some general properties of the solution of the DKP equation in a central interaction for spin zero particle. The central interaction consists of two parts: a Lorentz scalar $U_S$ and a time-like component of four-dimensional vector potential, $U_V$ depending only on $r$ [19,22]. The stationary states of the DKP particle, in units where $\hbar=c=1$, in this case are determined by solving

$$\left(\vec{\beta}.\vec{p}+m+U_S+\beta^0U_V\right)\psi(\vec{r})=\beta^0E\psi(\vec{r}). \tag{7}$$

In the spin zero representation, the five component DKP spinor:

$$\psi(\vec{r})=\begin{pmatrix}\psi_{\text{upper}}\\ i\psi_{\text{lower}}\end{pmatrix}\ \text{with}\ \psi_{\text{upper}}=\begin{pmatrix}\phi\\ \varphi\end{pmatrix}\ \text{and}\ \psi_{\text{lower}}=\begin{pmatrix}A_1\\ A_2\\ A_3\end{pmatrix}, \tag{8}$$

so that for stationary states the DKP equation can be written as

$$(m+U_S)\phi=(E-U_V^0)\varphi+\vec{\nabla}.\vec{A}, \tag{9}$$

$$\vec{\nabla}\phi=(m+U_S)\vec{A}, \tag{10}$$

$$(m+U_S)\varphi=(E-U_V^0)\phi, \tag{11}$$

where $\vec{A}$ is the vector $(A_1,A_2,A_3)$.The five-component wavefunction $\psi$ is simultaneously an eigenfunction of $J^2$ and $J_3$

$$J^2\begin{pmatrix}\psi_{\text{upper}}\\ \psi_{\text{lower}}\end{pmatrix}=\begin{pmatrix}L^2\psi_{\text{upper}}\\ (L+S)\psi_{\text{lower}}\end{pmatrix}=J(J+1)\begin{pmatrix}\psi_{\text{upper}}\\ \psi_{\text{lower}}\end{pmatrix}, \tag{12}$$

$$J_3\begin{pmatrix}\psi_{\text{upper}}\\ \psi_{\text{lower}}\end{pmatrix}=\begin{pmatrix}L_3\psi_{\text{upper}}\\ (L_3+S_3)\psi_{\text{lower}}\end{pmatrix}=M\begin{pmatrix}\psi_{\text{upper}}\\ \psi_{\text{lower}}\end{pmatrix}, \tag{13}$$

where the total angular momentum $J=L+S$ which commutes with $\beta^0$, is a constant of the motion. The most general solution of (7) is

$$\psi_{JM}(r)=\begin{pmatrix}f_{nJ}(r)Y_{JM}(\Omega)\\ g_{nJ}(r)Y_{JM}(\Omega)\\ i\sum_L f_{nJL}(r)Y_{JL1}^M(\Omega)\end{pmatrix}, \tag{14}$$

where $Y_{JM}(\Omega)$ are the spherical harmonics of order $J$, $Y_{JL1}^{M}(\Omega)$ are the normalized vector spherical harmonics and $f_{nJ}(r)$, $g_{nJ}(r)$ and $f_{nJL}(r)$ are radial wave functions. The insertion of $\psi_{JM}(r)$ given in Eq. (14) into Eqs. (9), (10) and (11) by using the properties of vector spherical harmonics [2] gives the following set of first-order coupled relativistic differential radial equations

$$(E-U_V^0)F(r)=(m+U_S)G(r), \tag{15a}$$

$$\left(\frac{d}{dr}-\frac{J+1}{r}\right)F(r)=-\frac{1}{\alpha_J}(m+U_S)H_1(r), \tag{15b}$$

$$\left(\frac{d}{dr}+\frac{J}{r}\right)F(r)=-\frac{1}{\varsigma_J}(m+U_S)H_{-1}(r), \tag{15c}$$

$$\begin{aligned}-\alpha_J\left(\frac{d}{dr}+\frac{J+1}{r}\right)H_1(r)+\varsigma\left(\frac{d}{dr}-\frac{J}{r}\right)H_{-1}(r)\\ =(m+U_S)F(r)-(E-U_V^0)G(r),\end{aligned} \tag{15d}$$

where $\alpha_J=\sqrt{(J+1)/(2J+1)}$, $\varsigma=\sqrt{J/(2J+1)}$, $f_{nJ}(r)=F(r)/r$, $g_{nJ}(r)=G(r)/r$ and $h_{nJJ\pm1}(r)=H_{\pm1}(r)/r$. For DKP equation, at the presence of vector potential and while scalar potential is zero, the differential equations to be satisfied by the radial wavefunctions are

$$(E-U_V^0)F(r)=mG(r), \tag{16a}$$

$$\left(\frac{d}{dr}-\frac{J+1}{r}\right)F(r)=-\frac{1}{\alpha_J}mH_1(r), \tag{16b}$$

$$\left(\frac{d}{dr}+\frac{J}{r}\right)F(r)=-\frac{1}{\varsigma_J}mH_{-1}(r), \tag{16c}$$

$$-\alpha_J\left(\frac{d}{dr}+\frac{J+1}{r}\right)H_1(r)+\varsigma\left(\frac{d}{dr}-\frac{J}{r}\right)H_{-1}(r)$$
$$=mF(r)-(E-U_V^0)G(r). \qquad (16d)$$

Eliminating $G(r)$, $H_1(r)$ and $H_{-1}(r)$ in terms of $F(r)$, the following second-order differential equation is satisfied by the function $F(r)$,

$$\vec{O}_{\text{KG}}F(r)=0, \qquad (17a)$$

which is the radial KG equation for a vector potential with $\vec{O}_{\text{KG}}$ is the Klein-Gordon operator defined as

$$\vec{O}_{\text{KG}}\rightarrow\vec{\nabla}^2+\left(E-eV(r)\right)^2-m^2,\ \vec{\nabla}^2\rightarrow\frac{d^2}{dr^2}-\frac{J(J+1)}{r^2}.$$

It is remarkable to note that Eq. (17a) is equivalent to Eq. (16) of Ref. [17]. Alternatively, we can rewrite (17a) as the radial Klein-Gordon equation [19]:

$$\left[\frac{d^2}{dr^2}-\frac{J(J+1)}{r^2}+\left(E-U_V^0\right)^2-m^2\right]F(r)=0,\ U_V^0=eV(r), \qquad (17b)$$

where $J(J+1)/r^2$, is the total angular momentum centrifugal term.

Here, we will consider an important inter-nuclear potential that represents the interaction between the projectile and the target nuclei in the form of the Woods-Saxon (WS) potential [24,25,34]. It plays an essential role in nuclear physics and microscopic physics, since it can be used to describe the nucleon-heavy-nucleus interactions. Therefore, we take the vector potential in Eq. (17b) as the dWS potential given by [35]

$$U_V^0(r)=\frac{U_0}{1+q\,\mathrm{e}^{(r-R)/a}},\quad R=r_0A_0^{1/3},\quad U_0=(40.5+0.13A_0)\,MeV\,,\ R\gg a,\ q>0 \qquad (18)$$

where $U_0$ is the depth of potential, $q$ is the real parameter which determine the shape (deformation) of the potential, $a$ is the diffuseness of the nuclear surface, $R$ is the width of the potential, $A_0$ is the atomic mass number of target nucleus and $r_0$ is radius parameter. After inserting (18) into (17b), the radial DKP equation for $F(r)$ reduces to

$$\left[\frac{d^2}{dr^2}-\frac{J(J+1)}{r^2}+\frac{U_0^2}{\left(1+q\,\mathrm{e}^{(r-R)/a}\right)^2}-\frac{2EU_0}{1+q\,\mathrm{e}^{(r-R)/a}}+E^2-m^2\right]F(r)=0. \qquad (19)$$

### 2.1. Pekeris-type approximation to the centrifugal term

Because of the total angular momentum centrifugal term, Eq. (19) cannot be solved analytically for $J \neq 0$. Therefore, we shall use the Pekeris approximation [36-38] in order to deal with this centrifugal term and we may express it as follows

$$U_{\text{centrifugal}} = \frac{J(J+1)}{r^2} = \frac{J(J+1)}{R^2\left(1+\frac{x}{R}\right)^2} \cong \frac{J(J+1)}{R^2}\left(1-2\frac{x}{R}+3\left(\frac{x}{R}\right)^2+\ldots\right), \tag{20}$$

with $x = r - R$. In addition, we may also approximately express it in the following way

$$\tilde{U}_{\text{centrifugal}} = \frac{J(J+1)}{r^2} \cong \frac{J(J+1)}{R^2}\left[D_0 + \frac{D_1}{1+qe^{\nu x}} + \frac{D_2}{\left(1+qe^{\nu x}\right)^2}\right], \tag{21}$$

where $\nu = 1/a$. After expanding (21) in terms of $x$ , $x^2$, $x^3$, … and next, comparing with Eq. (20), we obtain expansion coefficients $D_0$, $D_1$ and $D_2$ as follows

$$D_0 = 1 - \frac{(1+q)^2}{\nu R q^2}\left(1-\frac{3}{\nu R}\right),$$
$$D_1 = \frac{(1+q)^2}{\nu R q^2}\left(-1+3q-\frac{6(1+q)}{\nu R}\right),$$
$$D_2 = \frac{(1+q)^3}{\nu R q^2}\left(\frac{1-q}{2}+\frac{3(1+q)}{\nu R}\right). \tag{22}$$

Now, we can take the potential $\tilde{U}_{\text{centrifugal}}$ (21) instead of the total angular momentum centrifugal potential (20). By substituting (21) into (19), we obtain

$$\left[\frac{d^2}{dx^2} - \frac{J(J+1)}{R^2}\left(D_0 + \frac{D_1}{1+qe^{\nu x}} + \frac{D_2}{\left(1+qe^{\nu x}\right)^2}\right) + \frac{U_0^2}{\left(1+q\,\mathrm{e}^{\nu x}\right)^2} - \frac{2EU_0}{1+q\,\mathrm{e}^{\nu x}} + E^2 - m^2\right]F(x) = 0. \tag{23}$$

In the next section, we will introduce the generalized parametric NU method so that we can find solutions to the above equation.

## 3. Nikiforov-Uvarov method

To solve second order differential equations, the NU method can be used with an appropriate coordinate transformation $s = s(r)$ [39]

$$\psi_n''(s) + \frac{\tilde{\tau}(s)}{\sigma(s)}\psi_n'(s) + \frac{\tilde{\sigma}(s)}{\sigma^2(s)}\psi_n(s) = 0, \tag{24}$$

where $\sigma(s)$ and $\tilde{\sigma}(s)$ are polynomials, at most of second-degree, and $\tilde{\tau}(s)$ is a first-degree polynomial. The following equation is a general form of the Schrödinger-like equation written for any potential [40]

$$\left[\frac{d^2}{ds^2} + \frac{\alpha_1 - \alpha_2 s}{s(1-\alpha_3 s)}\frac{d}{ds} + \frac{-\xi_1 s^2 + \xi_2 s - \xi_3}{\left[s(1-\alpha_3 s)\right]^2}\right]\psi_n(s) = 0. \tag{25}$$

According to the NU method, the eigenfunctions and the energy eigenvalue equation are found as

$$\psi(s) = s^{\alpha_{12}}(1-\alpha_3 s)^{-\alpha_{12} - \frac{\alpha_{13}}{\alpha_3}} P_n^{\left(\alpha_{10}-1, \frac{\alpha_{11}}{\alpha_3} - \alpha_{10} - 1\right)}(1 - 2\alpha_3 s), \tag{26}$$

$$\alpha_2 n - (2n+1)\alpha_5 + (2n+1)\left(\sqrt{\alpha_9} + \alpha_3\sqrt{\alpha_8}\right) + n(n-1)\alpha_3 + \alpha_7 + 2\alpha_3\alpha_8 + 2\sqrt{\alpha_8\alpha_9} = 0, \tag{27}$$

respectively, with analytical parameters

$$\begin{aligned} \alpha_4 &= \frac{1}{2}(1-\alpha_1), & \alpha_5 &= \frac{1}{2}(\alpha_2 - 2\alpha_3), \\ \alpha_6 &= \alpha_5^2 + \xi_1, & \alpha_7 &= 2\alpha_4\alpha_5 - \xi_2, \\ \alpha_8 &= \alpha_4^2 + \xi_3 & \alpha_9 &= \alpha_6 + \alpha_3\alpha_7 + \alpha_3^2\alpha_8, \end{aligned} \tag{28}$$

and

$$\begin{aligned} \alpha_{10} &= \alpha_1 + 2\alpha_4 + 2\sqrt{\alpha_8}, & \alpha_{11} &= \alpha_2 - 2\alpha_5 + 2\left(\sqrt{\alpha_9} + \alpha_3\sqrt{\alpha_8}\right), \\ \alpha_{12} &= \alpha_4 + \sqrt{\alpha_8}, & \alpha_{13} &= \alpha_5 - \left(\sqrt{\alpha_9} + \alpha_3\sqrt{\alpha_8}\right). \end{aligned} \tag{29}$$

In some cases, we may have $\alpha_3 = 0$. Under this limit, the Jacobi polynomials turn to the Laguerre polynomials as

$$\lim_{\alpha_3 \to 0} P_n^{\left(\alpha_{10}-1, \frac{\alpha_{11}}{\alpha_3} - \alpha_{10} - 1\right)}(1-\alpha_3)s = L_n^{\alpha_{10}-1}(\alpha_{11} s), \tag{30}$$

and

$$\lim_{\alpha_3\to 0}\left(1-\alpha_3 s\right)^{-\alpha_{12}-\frac{\alpha_{13}}{\alpha_3}}=e^{\alpha_{13}s}, \qquad (31)$$

the solution given in Eq. (26) becomes as [40]

$$\psi(s)=s^{\alpha_{12}}e^{\alpha_{13}s}L_n^{\alpha_{10}-1}\left(\alpha_{11}s\right). \qquad (32)$$

## 4. Solution of the DKP equation with a vector deformed Woods-Saxon potential

Now we will solve Eq. (23) by using the parametric generalization NU method. Thus, we introduce change of variables $s=\frac{1}{1+qe^{vx}}$, which maps the interval $(0,\infty)$ into $(0,1)$, to rewrite it as follows:

$$\frac{d^2F(s)}{ds^2}+\frac{1-2s}{s(1-s)}\frac{dF(s)}{ds}+\frac{a^2}{s^2(1-s)^2}\Big[-\frac{J(J+1)}{R^2}(D_0+D_1 s+D_2 s^2)$$
$$-2EU_0 s+U_0^2 s^2+E^2-m^2\Big]F(s)=0. \qquad (33)$$

Comparing Eq. (33) with its counterpart Eq. (25), we can easily obtain the coefficients $\alpha_i$ ($i=1,2,3$) and analytical expressions $\xi_j$ ($j=1,2,3$) as follows

$$\alpha_1=1, \qquad \xi_1=a^2\left(\frac{J(J+1)}{R^2}D_2-U_0^2\right),$$
$$\alpha_2=2, \qquad \xi_2=-a^2\left(\frac{J(J+1)}{R^2}D_1+2EU_0\right),$$
$$\alpha_3=1, \qquad \xi_3=a^2\left(\frac{J(J+1)}{R^2}D_0+m^2-E^2\right). \qquad (34)$$

The remaining values of coefficients $\alpha_i$ ($i=4,5,...,13$) are found from relations (28) and (29). All values of these coefficients, i.e., $\alpha_i$ ($i=1,2,...,13$) together with $\xi_j$ ($j=1,2,3$) are displayed in table 1. By using Eq. (27), we can obtain, in closed form, the energy eigenvalue equation as

$$\left[n+\frac{1}{2}+a\left(\sqrt{\frac{J(J+1)}{R^2}(D_0+D_1+D_2)+m^2-(E_{nJ}-U_0)^2}+\sqrt{\frac{J(J+1)D_0}{R^2}+m^2-E_{nJ}^2}\right)\right]^2$$
$$=\frac{1}{4}+a^2\left(\frac{J(J+1)D_2}{R^2}-U_0^2\right). \qquad (35)$$

To find the energy levels we solve Eq. (35) numerically considering the following values of the parameters for $^{208}Pb$ as $m = 938 MeV$, $U_0 = -67.54 MeV$, $R = 7.6136 fm$, $a = 0.65 fm^{-1}$ [41,42] and $q = 1.5$. These numerical results are displayed in table 2. Further, in tables 3 and 4, we present the bound state energy eigenvalues for various shape of the potential parameter, i.e. $q$ and for various width of the potential, i.e. $R$, respectively.

To find corresponding wave functions, referring to table 1 and relation (26), we get

$$F(r) = N_{nJ}\left(\frac{1}{1+qe^{(r-R)/a}}\right)^{w_1}\left(1-\frac{1}{\left(1+qe^{(r-R)/a}\right)}\right)^{w_2} P_n^{(2w_1,2w_2)}\left(1-\frac{2}{\left(1+qe^{(r-R)/a}\right)}\right), \tag{36a}$$

with

$$w_1 = a\sqrt{\left(J(J+1)D_0/R^2 + m^2 - E_{nJ}^2\right)}, \tag{36b}$$

$$w_2 = a\sqrt{J(J+1)(D_0+D_1+D_2)/R^2 + m^2 - (E_{nJ}-U_0)^2}, \tag{36c}$$

and $N_{nJ}$ is the normalization constant. The wave function (36a) satisfies the standard asymptotic analysis for $r \to 0$ and $r \to \infty$. We can take into account the following hypergeometric property [43,44]

$$\frac{d}{dx}\left({}_2F_1(b,c;d;x)\right) = \left(\frac{bc}{d}\right){}_2F_1(b+1,c+1;d+1;x), \tag{37}$$

and express the hypergemetric function in terms of the Jacobi polynomials [42,43]

$$P_n^{(\lambda,\eta)}(1-2x) = \frac{(\lambda+1)_n}{n!}\,{}_2F_1(-n,\lambda+\eta+1+n;\lambda+1;x), \tag{38}$$

where $(y)_n = \Gamma(y+1)/\Gamma(y-n+1)$, is the Pochhammer's symbol. Thereby, we seek to find the other spinor components from Eqs. (16a), (16b) and (16c) as [44]

$$G(r) = N_{nJ}\frac{(2w_1+1)_n}{n!}\frac{1}{m}\left(E_{nJ} - \frac{U_0}{1+q\,\mathrm{e}^{(r-R)/a}}\right)\left(\frac{1}{1+qe^{(r-R)/a}}\right)^{w_1}\left(1-\frac{1}{\left(1+qe^{(r-R)/a}\right)}\right)^{w_2}$$

$$\times{}_2F_1\left(-n, 2(w_1+w_2)+1+n, 2w_1+1; \frac{1}{1+qe^{(r-R)/a}}\right), \tag{39a}$$

$$H_1(r) = -\frac{1}{m}\sqrt{\frac{J+1}{2J+1}}\left[\frac{1}{a}\left(w_2 - w_1qe^{(r-R)/a}\right)\frac{1}{1+qe^{(r-R)/a}} - \frac{J+1}{r}\right]F(r)$$

$$-N_{nJ}\frac{nq}{am}\sqrt{\frac{J+1}{2J+1}}\frac{(2w_1+2w_2+n+1)}{(2w_1+1)}\frac{(2w_1+1)_n}{n!}e^{(r-R)/a}\left(\frac{1}{1+qe^{(r-R)/a}}\right)^{w_1+2}$$

$$\times\left(1-\frac{1}{1+qe^{(r-R)/a}}\right)^{w_2}{}_2F_1\left(1-n,2(w_1+w_2+1)+n,2(w_1+1);\frac{1}{1+qe^{(r-R)/a}}\right), \tag{39b}$$

$$H_{-1}(r)=-\frac{1}{m}\sqrt{\frac{J}{2J+1}}\left[\frac{1}{a}\left(w_2-w_1qe^{(r-R)/a}\right)\frac{1}{1+qe^{(r-R)/a}}+\frac{J}{r}\right]F(r)$$

$$-N_{nJ}\frac{nq}{am}\sqrt{\frac{J}{2J+1}}\frac{(2w_1+2w_2+n+1)}{(2w_1+1)}\frac{(2w_1+1)_n}{n!}e^{(r-R)/a}\left(\frac{1}{1+qe^{(r-R)/a}}\right)^{w_1+2}$$

$$\times\left(1-\frac{1}{1+qe^{(r-R)/a}}\right)^{w_2}{}_2F_1\left(1-n,2(w_1+w_2+1)+n,2(w_1+1);\frac{1}{1+qe^{(r-R)/a}}\right). \tag{39c}$$

## 4. $J=0$ Case

We consider the special case when the total angular momentum $J=0$ (s-wave). Thus, from Eq. (35) we obtain the following exact energy equation:

$$1+2n+2a\left(\sqrt{m^2-(E_{n0}-U_0)^2}+\sqrt{m^2-E_{n0}^2}\right)=\sqrt{1-4a^2U_0^2}, \tag{40}$$

where $|m|>(E_{n0}-U_0)$, $|m|>E_{n0}$ and $1>2aU_0$. In case if we follow the same notations and procedure of solution presented by Ref. [45], we can obtain the energy formula (10) for the S-wave. However, their solution is only valid for the case $R=0$ as demonstrated in [46,47]. Further, from Eqs. (36) and (39), we also find the exact spinor components of the wave function as

$$F(r)=N_n\left(\frac{1}{1+qe^{(r-R)/a}}\right)^{p_1}\left(1-\frac{1}{1+qe^{(r-R)/a}}\right)^{p_2}P_n^{(2p_1,2p_2)}\left(1-\frac{2}{1+qe^{(r-R)/a}}\right), \tag{41a}$$

$$G(r)=N_n\frac{(2p_1+1)_n}{n!}\frac{1}{m}\left(E_{n0}-\frac{U_0}{1+q\,\mathrm{e}^{(r-R)/a}}\right)\left(\frac{1}{1+qe^{(r-R)/a}}\right)^{p_1}\left(1-\frac{1}{1+qe^{(r-R)/a}}\right)^{p_2}$$

$$\times{}_2F_1\left(-n,2(p_1+p_2)+1+n,2p_1+1;\frac{1}{1+qe^{(r-R)/a}}\right), \tag{41b}$$

$$H_1(r)=-\frac{1}{m}\left[\frac{1}{a}\left(p_2-p_1qe^{(r-R)/a}\right)\frac{1}{1+qe^{(r-R)/a}}-\frac{1}{r}\right]F(r)$$

$$-N_n\frac{nq}{am}\frac{(2p_1+2p_2+n+1)}{(2p_1+1)}\frac{(2p_1+1)_n}{n!}e^{(r-R)/a}\left(\frac{1}{1+qe^{(r-R)/a}}\right)^{p_1+2}$$

$$\times\left(1-\frac{1}{1+qe^{(r-R)/a}}\right)^{p_2}\ {}_2F_1\left(1-n,2(p_1+p_2+1)+n,2(p_1+1);\frac{1}{1+qe^{(r-R)/a}}\right), \tag{41c}$$

$$H_{-1}(r)=0, \tag{41d}$$

with

$$p_1=a\sqrt{m^2-E_{n0}^2},\ p_2=a\sqrt{m^2-(E_{n0}-U_0)^2},\ U_0=eV_0. \tag{42}$$

## 5. Conclusion

In this paper, we solved the Duffin–Kemmer–Petiau equation under a vector dWS potential. We used the Pekeris approximation to the total angular momentum centrifugal term. The parametric NU method is used to obtain the energy eigenvalues and the corresponding eigenfunctions. The explicit forms of the spinor components of wave function were calculated. In addition, some numerical results for bound state energies are given in tables 2, 3 and 4 for various $(n,J)$ states by considering different nuclear deformations $q$ and widths $R$ for the dWS interactions.

## Acknowledgments

The authors wish to thank the kind referees for their invaluable suggestions which have greatly helped in improving this paper.

**Table 1.** The specific values for the parametric constants necessary for the energy eigenvalues and eigenfunctions

| Constant | Analytic value |
|---|---|
| $\alpha_1$ | 1 |
| $\alpha_2$ | 2 |
| $\alpha_3$ | 1 |
| $\alpha_4$ | 0 |
| $\alpha_5$ | 0 |
| $\alpha_6$ | $a^2\left(\frac{J(J+1)D_2}{R^2}-U_0^2\right)$ |
| $\alpha_7$ | $a^2\left(\frac{J(J+1)D_1}{R^2}+2EU_0\right)$ |
| $\alpha_8$ | $a^2\left(\frac{J(J+1)D_0}{R^2}+m^2-E^2\right)$ |
| $\alpha_9$ | $\frac{J(J+1)a^2}{R^2}(D_0+D_1+D_2)+m^2-a^2\left((E-U_0)^2\right)$ |
| $\alpha_{10}$ | $1+2a\sqrt{\frac{J(J+1)D_0}{R^2}+m^2-E^2}$ |
| $\alpha_{11}$ | $2+2a\left(\sqrt{\frac{J(J+1)}{R^2}(D_0+D_1+D_2)+m^2-\left((E-U_0)^2\right)}\right.$ |
| | $\left.+\sqrt{\frac{J(J+1)D_0}{R^2}+m^2-E^2}\right)$ |
| $\alpha_{12}$ | $a\sqrt{\frac{J(J+1)D_0}{R^2}+m^2-E^2}$ |
| $\alpha_{13}$ | $-a\left(\sqrt{\frac{J(J+1)}{R^2}(D_0+D_1+D_2)+m^2-\left((E-U_0)^2\right)}\right.$ |
| | $\left.+\sqrt{\frac{J(J+1)D_0}{R^2}+m^2-E^2}\right)$ |
| $\xi_1$ | $a^2\left(\frac{J(J+1)D_2}{R^2}-U_0^2\right)$ |
| $\xi_2$ | $-a^2\left(\frac{J(J+1)D_1}{R^2}+2EU_0\right)$ |
| $\xi_3$ | $a^2\left(\frac{J(J+1)D_0}{R^2}+m^2-E^2\right)$ |

**Table 2:** Bound state eigenvalues $E_{nJ}\,(MeV)$ of the DKP equation under a vector deformed Woods-Saxon potential for various $n$ and $J$ 's.

| $n$ | $J$ | $E_{n,J}$ |
|---|---|---|
| 0 | 0 | -935.4762526 |
| | 1 | -936.0290545 |
| | 2 | -937.1322307 |
| 1 | 0 | -917.7895470 |
| | 1 | -918.4243638 |
| | 2 | -919.6947876 |
| 2 | 0 | -855.7387858 |
| | 1 | -856.3670262 |
| | 2 | -857.6275738 |
| 3 | 0 | -754.7038394 |
| | 1 | -755.3233006 |
| | 2 | -756.5701831 |

**Table 3:** Bound state eigenvalues $E_{nJ}\,(MeV)$ of the DKP equation under a vector deformed Woods-Saxon potential for various $q$ s.

| $q$ | $E_{0,1}$ | $E_{1,1}$ | $E_{2,1}$ | $E_{2,2}$ |
|---|---|---|---|---|
| 1 | -936.0388248 | -918.2641628 | -856.1251707 | -856.9111293 |
| 1.1 | -936.0384065 | -918.3071211 | -856.1893642 | -857.1010868 |
| 1.2 | -936.0370346 | -918.3429337 | -856.2431768 | -857.2604400 |
| 1.3 | -936.0349274 | -918.3735850 | -856.2895136 | -857.3977374 |
| 1.4 | -936.0322327 | -918.4004167 | -856.3303394 | -857.5187667 |
| 1.5 | -936.0290545 | -918.4243638 | -856.3670262 | -857.6275738 |

**Table 4:** Bound state eigenvalues $E_{nJ}\,(fm^{-1})$ of the DKP equation under a vector deformed Woods-Saxon potential for various $R$ s.

| $R$ | $E_{0,1}$ | $E_{1,1}$ | $E_{2,1}$ | $E_{2,2}$ |
|---|---|---|---|---|
| 6 | -936.4080762 | -918.6526276 | -856.4484175 | -857.9145772 |
| 7 | -936.1367961 | -918.5059640 | -856.4168376 | -857.7834909 |
| 8 | -935.9752972 | -918.3779451 | -856.3329411 | -857.5233423 |
| 9 | -935.8699722 | -918.2753520 | -856.2470283 | -857.2635451 |